\documentclass[
reprint,
aps,
prd,
superscriptaddress,
nofootinbib
]{revtex4-2}

\usepackage[T1]{fontenc}
\usepackage[utf8]{inputenc}
\usepackage{amsmath,amssymb,amsthm,mathtools}
\usepackage{bm}
\usepackage{hyperref}
\usepackage{tikz}
\usetikzlibrary{arrows.meta}
\usetikzlibrary{arrows.meta,calc}

\begin{document}

\title{Electromagnetic Field Equivalence from Moving Manifolds}

%\title{Electromagnetic Field Matching from Moving-Manifold Dynamics}

\author{David V. Svintradze}
\affiliation{New Vision University, Tbilisi, Georgia}

\date{\today}

\begin{abstract}
We apply a geometric formulation of electromagnetic fields on moving manifolds to the problem of field equivalence between dynamically separated domains. Starting from the tensorially invariant equations of motion for moving hypersurfaces, we introduce an electromagnetic specialization by constructing an energy density from the electromagnetic field tensor, yielding a geometric extension of Maxwell electrodynamics. The classical Maxwell equations then emerge as a constrained geometric sector of the broader evolution system. Hence, by comparing internal and external electromagnetic configurations, we show that under isolation conditions with no interfacial current exchange, the field difference satisfies the source-free Maxwell equation. 
Furthermore, the equilibrium Maxwell sector establishes a direct correspondence between the Lorentz-invariant electromagnetic structure and the geometry of constant-mean-curvature manifolds.
 The resulting field-difference equations admit equivalence solutions generated by specific velocity sectors of the moving-manifold dynamics. We further demonstrate that the resulting field-equivalence regime is intrinsically dynamical: static curved configurations generically retain nonvanishing electromagnetic contrast through curvature-induced contributions to the geometric pressure balance, whereas dynamically evolving manifolds admit nontrivial admissible sectors satisfying the equivalence condition. Explicit nonvacuum realizations are obtained within the tangential-flow sector of the moving-manifold system, while bounded static Euclidean configurations are generically excluded.
\end{abstract}

\maketitle

\section{introduction}

Electromagnetic cloaking has been studied extensively using transformation optics and conformal mappings, in which the spatial variation of permittivity and permeability allows the control of electromagnetic field paths and their restoration to the original state \cite{Pendry2006,Leonhardt2006}. Experimental observations have shown effectiveness in microwave, optical and visible regimes
\cite{Schurig2006,Liu2009,Tretyakov2009,Edwards2009,Ma2010,Valentine2009,Gabrielli2009,Smolyaninov2009,Ergin2010,Chen2011}. Nevertheless, such approaches are limited, namely, they depend on photon wavelength and scales of material %\cite{Cho2010}
, do not take into account time variations, and are highly dependent on the shape of the object. Theoretical physics has intensively studied the relationship between Geometry and EM fields. In particular, the local conservation laws are incorporated into Maxwell’s equations that describe all electromagnetic phenomena \cite{Vanderlinde2010}. However, these equations characterize fields generated by specified sources on a fixed domain, assuming that the charge or current density remains stationary. The moving boundary brings new geometric structures to electromagnetism and, therefore, couples Maxwell’s equations to the process of domain evolution. In other words, writing Maxwell’s equations on fixed domains turns out to be insufficient for modeling such problems \cite{Svintradze2017}.

We adopt a different approach from transformation optics. Namely, we study EM fields using the moving manifold (MM) framework, in which the evolving shape of a hypersurface enters the equations of motion, thereby integrating geometric considerations into the governing equations. We begin from a geometric evolution system that integrates curvature, normal motion, and tangential flow. By introducing an energy density derived from the electromagnetic field tensor, the governing equations assume an extended geometric form wherein deviations from Maxwell’s equations are represented by a kinetic energy-like structure induced by manifold dynamics. Hence, the action takes the following form:
\begin{align}
\mathcal{L}
&=
\int_{\Sigma(t)} \frac{\rho V^2}{2}\, d\Sigma
-
\int_{\Sigma(t)} \sigma d\Sigma
-
\int_{\Omega(t)} P d\Omega \label{Lagrangian} \\
P
&=
\frac{1}{4\mu_0}F_{ab}F^{ab}
-
A_aJ^a \label{EM_pressure}
\end{align}
where $P$ is the energy density (in this work, we assume $P$ to be the EM energy density), $F_{ab}$ is the electromagnetic field tensor, $A_a$ is the four-vector potential, $J^a$ is the four-current, and Latin letters stand for ambient flat Minkowski space tensor indices $a=0, 1, ..., 3$. The quantity $\rho$ denotes the hypersurface mass density or density field in general, while $\sigma$ represents the surface energy density associated with the evolving manifold \cite{Svintradze2026}. The domain $\Omega(t)$ is the ambient Minkowski spacetime bounded by the hypersurface $\Sigma(t)$, where $t$ is a parameter describing the evolution of the manifold rather than the physical proper time \cite{Svintradze2017}. Notably, the present formulation is generally applicable to arbitrary energy density fields $P$ and yields equations of motion for moving manifolds in a tensorially invariant form \cite{Svintradze2017,Svintradze2018,Svintradze2019,Svintradze2020,Svintradze2023,Svintradze2024}. Details about moving-manifold calculus and its broader theoretical structure are presented in \cite{Svintradze2025,Svintradze2026}. For related formulations of calculus on moving surfaces, see also \cite{Grinfeld2013}.

Given that the electromagnetic response is no longer solely determined by the source distribution but also by the geometry and dynamics of the supporting domain, this framework facilitates a direct comparison between internal and external electromagnetic configurations. Consequently, the equilibrium massless sector of the moving-manifold system reduces to a constant-mean-curvature (CMC) geometry, establishing a direct correspondence between Lorentz-invariant electromagnetic structure and equilibrium manifold CMC shapes. Consequently, Maxwell EM acquires a geometric interpretation through equilibrium CMC manifolds. The difference between these configurations satisfies a source-free Maxwell equation under isolation conditions. In the absence of permissible free electromagnetic modes, this condition entails the equality of internal and external fields. The resulting field-equivalence regime is inherently dynamical. In static configurations, curvature induces a nonzero electromagnetic contrast, whereas the dynamics of moving manifolds admit regimes in which this contrast vanishes. Explicit solutions in the flat case demonstrate that the class of such configurations is nonempty. Consequently, a geometric mechanism emerges whereby electromagnetic fields propagate undisturbed across a bounded region, establishing an equivalence between internal and external field configurations. This mechanism is not exclusive to electromagnetic fields but arises from the structure of the governing equations and extends to any field described within the same moving manifold framework \cite{Svintradze2023,Svintradze2026}.

\section{Problem Formulation}
%\section{Geometric Formulation of the Problem}

%\subsection{Schematic Representation} 

\begin{figure}[!ht]
\centering
\begin{tikzpicture}[scale=1.05, line cap=round, line join=round]

% initial domain Omega(0)
\draw[gray!55, thick, dashed]
  (-1.65,0) .. controls (-1.55,0.95) and (-0.65,1.35) .. (0.15,1.15)
  .. controls (1.05,0.95) and (1.55,0.35) .. (1.35,-0.45)
  .. controls (1.10,-1.25) and (0.05,-1.45) .. (-0.85,-1.10)
  .. controls (-1.55,-0.82) and (-1.85,-0.45) .. (-1.65,0);

\node[gray!60!black] at (-1.35,1.15) {$\Sigma(0)$};
\node[gray!60!black] at (-0.35,0.05) {$\Omega(0)$};

% evolved domain Omega(t)
\fill[blue!6]
  (-1.05,0.15) .. controls (-1.05,1.05) and (-0.20,1.55) .. (0.75,1.30)
  .. controls (1.65,1.05) and (2.05,0.20) .. (1.75,-0.65)
  .. controls (1.42,-1.55) and (0.35,-1.65) .. (-0.55,-1.20)
  .. controls (-1.35,-0.82) and (-1.42,-0.25) .. (-1.05,0.15);

\draw[blue!60!black, thick]
  (-1.05,0.15) .. controls (-1.05,1.05) and (-0.20,1.55) .. (0.75,1.30)
  .. controls (1.65,1.05) and (2.05,0.20) .. (1.75,-0.65)
  .. controls (1.42,-1.55) and (0.35,-1.65) .. (-0.55,-1.20)
  .. controls (-1.35,-0.82) and (-1.42,-0.25) .. (-1.05,0.15);

\node[blue!60!black] at (1.65,1.12) {$\Sigma(t)$};
\node[blue!60!black] at (0.35,0.05) {$\Omega(t)$};

% motion arrows from old to new boundary
\draw[-{Latex[length=2mm]}, gray!70, thick]
  (-1.60,0.45) -- (-1.08,0.55);
\draw[-{Latex[length=2mm]}, gray!70, thick]
  (0.10,1.15) -- (0.55,1.33);
\draw[-{Latex[length=2mm]}, gray!70, thick]
  (1.25,-0.55) -- (1.70,-0.72);
\draw[-{Latex[length=2mm]}, gray!70, thick]
  (-0.65,-1.05) -- (-0.25,-1.30);

% internal and external energy labels
\node at (0.35,-0.45) {$\underset{\sim}{P}$};
\node at (2.35,0.10) {$\overset{\sim}{P}$};

% optional field-region labels
\node[gray!65] at (2.45,-0.32) {External};
\node[gray!65] at (0.35,-0.78) {Internal};

\end{tikzpicture}

\caption{
Schematic formulation of the problem. A bounded region $\Omega(t)$ evolves with boundary $\Sigma(t)$ from an initial configuration $\Omega(0)$ with boundary $\Sigma(0)$. The quantities $\underset{\sim}{P}$ and $\overset{\sim}{P}$ denote the internal and external energy densities, respectively. In the electromagnetic specialization considered here, $P=\frac{1}{4\mu_0}F_{ab}F^{ab}-A_aJ^a$. The field-equivalence problem is to determine conditions under which the internal and external configurations become equivalent under the motion of the boundary.
}
\label{fig:problem_setup}
\end{figure}
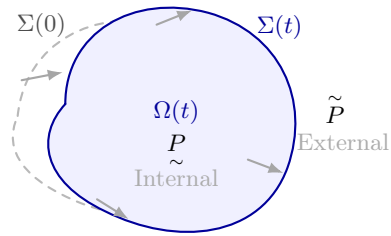

Electromagnetic fields are fundamentally defined on space-time domains and therefore naturally couple to the geometry of evolving boundaries separating distinct physical sectors, making four-dimensional Minkowski space-time $\Omega$ the natural ambient space for an evolving hypersurface $\Sigma(t)$. Here, the embedding is understood locally through the coordinate mapping $X^a=X^a(t,\xi^\alpha)$, where the manifold coordinates are generally complex-valued $\xi^\alpha\in\mathbb{C}$ in the pseudo-Riemannian Minkowski setting. No global embedding assumption beyond this local smooth parametrization is imposed. Consequently, within the moving-manifold formalism, the geometry of the boundary $\Sigma(t)$ is not passive but enters directly into the equations of motion through curvature, normal motion, and tangential flow. Although the ambient Minkowski space remains flat, the evolving hypersurface induces a time-dependent bounded domain $\Omega(t)$, where $t$ is a geometric evolution parameter not necessarily identified with physical proper time. Euclidean geometry then appears as a static limit of the more general pseudo-Riemannian setting. Hence, we consider a four-dimensional Minkowski spacetime $\Omega(t)$, which serves as the ambient space for the dynamics. Within this space, let $\Sigma(t)$ denote an evolving hypersurface that bounds a region of interest. The motion of $\Sigma(t)$ is described by a parametric representation $X^a = X^a(t,\xi^\alpha)$, where $a=0,1,2,3$ are ambient indices and $\alpha=0,1,2$ are hypersurface-associated tensors. A schematic representation of this setup is shown in Fig.~\ref{fig:problem_setup}. Throughout this work, $\Sigma(t)$ is assumed to be a smooth closed hypersurface unless stated otherwise.

Within the local moving-manifold parametrization, the evolving boundary naturally separates the surrounding space-time region into two dynamically coupled electromagnetic sectors: an internal configuration $\underset{\sim}{F}^{ab}$ associated with the bounded region enclosed by the hypersurface and an external configuration $\overset{\sim}{F}{}^{ab}$ defined in the surrounding domain. From the geometric point of view, the evolving hypersurface $\Sigma(t)$ acts as the interface through which the distinction between the internal and external sectors is established, while simultaneously constraining their admissible dynamics through curvature and boundary motion. More generally, within the moving-manifold framework, the quantity $P$ represents an arbitrary energy density field whose evolution is partitioned into internal and external contributions denoted by $\underset{\sim}{P}$ and $\overset{\sim}{P}$, respectively. Consequently, the formulation is not restricted to electromagnetic fields alone but applies to arbitrary physical systems whose dynamics admit a geometric energy-density representation. In the electromagnetic specialization considered throughout the present work, the energy density $P$ is specified by Eq.~\eqref{EM_pressure}, reducing the general MM framework to the problem of moving manifolds in electromagnetic fields. The central question is then to determine the geometric and dynamical conditions under which the internal and external electromagnetic configurations become equivalent despite the presence of a nontrivially evolving boundary separating the two sectors.

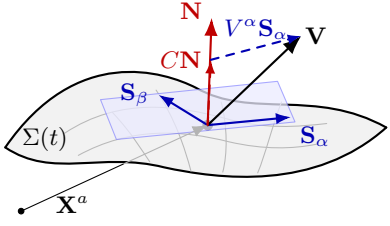
\begin{figure}[!t]
\centering
\begin{tikzpicture}[scale=1.05, line cap=round, line join=round]

% curved surface patch
\fill[gray!10]
  (-2.4,-0.6) .. controls (-1.5,0.7) and (-0.4,0.35) .. (0,0.15)
  .. controls (0.7,-0.25) and (1.5,0.15) .. (2.4,-0.35)
  .. controls (1.5,-1.15) and (0.3,-1.0) .. (-0.4,-0.85)
  .. controls (-1.2,-0.65) and (-1.8,-1.0) .. (-2.4,-0.6);

\draw[thick]
  (-2.4,-0.6) .. controls (-1.5,0.7) and (-0.4,0.35) .. (0,0.15)
  .. controls (0.7,-0.25) and (1.5,0.15) .. (2.4,-0.35);
\draw[thick]
  (2.4,-0.35) .. controls (1.5,-1.15) and (0.3,-1.0) .. (-0.4,-0.85)
  .. controls (-1.2,-0.65) and (-1.8,-1.0) .. (-2.4,-0.6);

% surface mesh
\draw[gray!55] (-1.65,-0.42) .. controls (-0.95,0.04) and (-0.22,-0.08) .. (0.35,-0.23)
  .. controls (0.88,-0.38) and (1.28,-0.34) .. (1.65,-0.30);
\draw[gray!55] (-1.48,-0.70) .. controls (-0.82,-0.34) and (-0.18,-0.38) .. (0.42,-0.53)
  .. controls (0.86,-0.63) and (1.22,-0.62) .. (1.52,-0.56);
\draw[gray!55] (-1.05,-0.16) .. controls (-0.78,-0.45) and (-0.68,-0.64) .. (-0.58,-0.78);
\draw[gray!55] (-0.50,0.12) .. controls (-0.26,-0.24) and (-0.14,-0.60) .. (-0.06,-0.90);
\draw[gray!55] (0.32,-0.03) .. controls (0.40,-0.34) and (0.40,-0.68) .. (0.34,-0.91);
\draw[gray!55] (1.08,-0.03) .. controls (0.96,-0.34) and (0.88,-0.68) .. (0.78,-0.92);

% point
\coordinate (P) at (0.15,-0.32);
\filldraw (P) circle (1.7pt);
%\node[below left] at ($(P)+(-0.15,-0.08)$) {$X(\xi,t)$};

% tangent plane, shifted visually behind point
\filldraw[fill=blue!8, draw=blue!45, opacity=0.75]
  ($(P)+(-1.15,-0.18)$) --
  ($(P)+(1.10,0.04)$) --
  ($(P)+(0.88,0.55)$) --
  ($(P)+(-1.35,0.32)$) -- cycle;

%\node[blue!60!black, anchor=east] at ($(P)+(-1.55,0.48)$) {$T_X\Sigma(t)$};

% tangent vectors
\draw[-{Latex[length=2mm]}, thick, blue!70!black]
  (P) -- ($(P)+(1.05,0.10)$) node[below right] {$\mathbf S_{\alpha}$};

\draw[-{Latex[length=2mm]}, thick, blue!70!black]
  (P) -- ($(P)+(-0.62,0.38)$) node[left] {$\mathbf S_{\beta}$};

% normal and velocity separated
\draw[-{Latex[length=2.4mm]}, thick, red!75!black]
  (P) -- ($(P)+(0.05,1.35)$) node[above left, yshift=-4pt] {$\mathbf N$};

\draw[-{Latex[length=2.5mm]}, thick, black]
  (P) -- ($(P)+(1.18,1.12)$) node[right, xshift=-2] {$\mathbf V$};

% components
\draw[-{Latex[length=1.8mm]}, dashed, thick, red!75!black]
  (P) -- ($(P)+(0.04,0.82)$) node[left] {$C\mathbf N$};

\draw[-{Latex[length=1.8mm]}, dashed, thick, blue!70!black]
  ($(P)+(0.04,0.82)$) -- ($(P)+(1.18,1.12)$)
  node[midway, above] {$V^\alpha\mathbf S_\alpha$};

% surface label
%\node at (1.95,0.35) {$\Sigma(t)$};
\node at (-1.9,-0.5) {$\Sigma(t)$};

% ambient origin and position vector X^A
\coordinate (O) at (-2.2,-1.4); % origin outside surface
\coordinate (Q) at ($(O)!0.55!(P)$); % transition point (entering surface)

% origin point
\filldraw (O) circle (1pt);

% outside segment (before surface)
\draw[-, thin]
  (O) -- (Q) node[midway, below] {$\mathbf X^a$};

% inside segment (faded, behind surface)
\draw[-{Latex[length=2mm]}, thin, gray!60]
  (Q) -- (P) ;

\end{tikzpicture}

\caption{
Local geometry of a moving manifold $\Sigma(t)$. The tangent directions are represented by $\mathbf S_\alpha$, while $\mathbf N$ denotes the unit normal. The velocity decomposes as $\mathbf V=C\mathbf N+V^\alpha\mathbf S_\alpha$, where $C$ is the normal velocity and $V^\alpha$ are tangential components. This decomposition is the kinematic basis of Eqs.~\eqref{eq:gen1}--\eqref{eq:gen3}. The vector $\mathbf X^{a}$ denotes the ambient position vector from the origin to the point on the manifold.
}
\label{fig:mm_geometry}
\end{figure}

%\subsection{Geometric Settings}
To keep the present work reasonably self-contained, we briefly summarize the minimal elements of moving-manifold calculus required for the subsequent analysis. A complete exposition of differential geometry and the calculus of moving surfaces is beyond the scope of the present paper and may be found in standard references such as \cite{Grinfeld2013,Sahraee2023}, together with the generalized moving-manifold formulation developed in \cite{Svintradze2017,Svintradze2025,Svintradze2026,Svintradze2026GMJ}. Within the local embedding framework introduced above, the evolving hypersurface $\Sigma(t)$ is described by its local differential geometry in the ambient Minkowski space-time. A smooth time-dependent hypersurface can therefore be represented by the position vector $\mathbf X^a(\xi^\alpha,t)$, where again $a=0,1,2,3$ denote ambient space-time indices and $\alpha=0,1,2$ are intrinsic coordinates on $\Sigma(t)$, as illustrated schematically in Fig.~\ref{fig:mm_geometry}. The local geometry of the hypersurface is determined by the tangent basis vectors
\[
\mathbf S_\alpha = \partial_\alpha \mathbf X,
\]
while the unit normal vector $\mathbf N$ satisfies the orthogonality and normalization conditions
\[
\mathbf N \cdot \mathbf S_\alpha = 0,
\qquad
\mathbf N \cdot \mathbf N = 1.
\]
The evolution of the hypersurface is characterized by the velocity field
\begin{equation}
\mathbf V = \partial_t \mathbf X = C\,\mathbf N + V^\alpha \mathbf S_\alpha, \label{velocity}
\end{equation}
which naturally decomposes into normal and tangential components. Here, $C$ denotes the normal velocity governing geometric deformation of the hypersurface, whereas $V^\alpha$ represents tangential velocity components associated with intrinsic surface flow. The extrinsic geometry of $\Sigma(t)$ is encoded through the curvature tensor $B_{\alpha\beta}$ defined by
\begin{equation}
\nabla_\alpha \mathbf S_\beta = B_{\alpha\beta}\,\mathbf N, \label{curvature}
\end{equation}
with mean curvature given by the trace $B_\alpha^{\ \alpha}$. Time evolution within the moving-manifold formalism is expressed through the invariant derivative 
\begin{equation}
\dot{\nabla}=\partial_t-V^\alpha\nabla_\alpha, \label{dotnabla}
\end{equation} 
which provides the covariant description of surface evolution in the presence of both normal and tangential motion. The notation introduced above represents only the minimal geometric structure required for the present work. A broader formulation of moving-manifold calculus, including invariant differential operators, fundamental geometric identities, integration theorems, and the full equations of motion, is developed in \cite{Svintradze2017,Svintradze2018,Svintradze2025,Svintradze2026,Svintradze2026GMJ,Grinfeld2013}.

\section{Geometric Electromagnetism and field equivalence} 

\subsection{Maxwell electrodynamics}
The field-equivalence problem formulated above cannot be addressed within a fixed-domain description, since the motion of the boundary modifies both the geometry of the manifold and the balance laws governing the fields supported on it. Consequently, the dynamics of the hypersurface itself must be incorporated into the governing equations. This requires a geometric evolution system in which curvature, normal motion, tangential flow, and field interactions are treated simultaneously within a tensorially invariant framework. To this end, we consider a smooth evolving manifold $\Sigma(t)$ whose motion is characterized by the normal velocity $C$ and tangential velocity components $V^{\alpha}$ \eqref{velocity}. The schematic representation of the velocity field on the $\Sigma(t)$ patch is given in Figure~\ref{fig:mm_geometry}. The geometry of the manifold is determined by the curvature tensor $B_{\alpha\beta}$ and its trace $B_{\alpha}^{\ \alpha}$ \eqref{curvature}, while temporal evolution is expressed through the invariant derivative $\dot{\nabla}$ \eqref{dotnabla}. The dynamics of a scalar density $\rho$ coupled to surface motion and curvature is governed by a closed geometric system derived from the stationary action principle associated with \eqref{Lagrangian}. The resulting equations represent an intrinsic extension of classical conservation and momentum laws to MM. The governing equations take the form
\begin{widetext}
\begin{align}
\dot{\nabla}\rho + \nabla_{\alpha}(\rho V^{\alpha})
&=
\rho C B_{\alpha}^{\ \alpha},
\label{eq:gen1}
\\
\partial_{a}V^{a}\Big(
\rho(\dot{\nabla}C
+ 2V^{\alpha}\nabla_{\alpha}C
+ V^{\alpha}V^{\beta}B_{\alpha\beta})
- \partial_t\sigma
- P
+ \sigma B_{\alpha}^{\ \alpha}
\Big)
&=
V^{a}\partial_{a}P,
\label{eq:gen2}
\\
\rho\Big(
\dot{\nabla}V^{\alpha}
+ V^{\beta}\nabla_{\beta}V^{\alpha}
- C\nabla^{\alpha}C
- C V^{\beta}B_{\beta}^{\ \alpha}
\Big)
&=
-\nabla^{\alpha}\sigma.
\label{eq:gen3}
\end{align}
\end{widetext}
Equation~\eqref{eq:gen1} expresses conservation of the surface density $\rho$ under motion and curvature of the manifold. Detailed derivations and the broader geometric framework are presented in \cite{Svintradze2017,Svintradze2018,Svintradze2023,Svintradze2025,Svintradze2026,Svintradze2026GMJ}. Equation~\eqref{eq:gen2} governs the normal dynamics of the hypersurface and shows the balance between geometric motion, surface energy, and volumetric forcing through the generalized energy density $P$. Equation~\eqref{eq:gen3} describes tangential deformation and represents the corresponding momentum balance along the manifold. The system is generic in that the form of $P$ is not specified a priori and may be constructed from any field that interacts with the manifold, thereby allowing different physical theories to be embedded within the same geometric framework.

To specialize the geometric evolution system to EM fields, we introduce the generalized energy density \eqref{EM_pressure}, where $F_{ab}=\partial_aA_b-\partial_b A_a$ is the electromagnetic field tensor, $A_a$ is the four-vector potential,  and $J^a$ is the four-current. The quantity $\sigma$ denotes the surface energy density associated with the moving manifold. Substitution of \eqref{EM_pressure} into the general MM system, considering that EM field lines cross the hypersurface perpendicularly, shows that the EM contribution affects only the normal balance \eqref{eq:gen2}. As a result, only the normal equation \eqref{eq:gen2} changes, leading to the coupled EM evolution equation: 
\begin{widetext}
\begin{align}
\mathcal{F}\mathcal{a}=\partial_{a}V^{a}\Big(
\rho(\dot{\nabla}C
+ 2V^{\alpha}\nabla_{\alpha}C
+ V^{\alpha}V^{\beta}B_{\alpha\beta})
-
\frac{1}{4\mu_0}F_{ab}F^{ab}
-
A_aJ^a
+
\sigma B_{\alpha}^{\ \alpha}
-
\partial_t\sigma
\Big),
\label{eq:em2}
\end{align}
\end{widetext}
where $\mathcal{A}=\bm A\cdot\bm N$ is the normal component of the four-vector potential $\bm A$ and
\begin{equation}
\mathcal{F}^{a}=J^{a}-\frac{1}{\mu_0}\partial_{b}F^{ba}.
\label{eq:residual}
\end{equation}
For details, see \cite{Svintradze2017}. Equation~\eqref{eq:em2} along with \eqref{eq:gen1}--\eqref{eq:gen3} describe the interplay between the dynamics of electromagnetic fields and moving manifolds. \(\mathcal {F} \) measures deviations from Maxwell EM inside the MM normal balance \eqref{eq:em2}, and we refer to it as the Maxwell field.
Note that the Maxwell equations emerge in the $C,\rho=0$ dynamic manifold conditions as a particular solution of \eqref{eq:em2} for time-independent $\sigma$ hypersurface fields:
\begin{equation}
\partial_{b}F^{ba}
=
\mu_0 J^{a},
\label{eq:maxwell}
\end{equation}
Thus, Maxwell's EM is a specific case of the MM equations \eqref{eq:gen1}--\eqref{eq:gen3} characterized by the specific velocity and density fields. Next, we study what condition induces the vanishing of \eqref{eq:residual} that naturally leads to \eqref{eq:maxwell} and therefore sets field equivalence in the case of two internal/external fields. 

Note that in the static massless equilibrium sector $C,\rho=0$ with constant surface energy density $\sigma$, the normal balance equation \eqref{eq:em2} admits the equilibrium solution
\begin{equation}
\frac{1}{4\mu_0}F_{ab}F^{ab}-A_aJ^a
=
\sigma B_\alpha^{\ \alpha} \label{CMC-Maxwell},
\end{equation}
provided that $\mathcal{F}=0$ and $P\neq 0$. Since the condition $\mathcal{F}=0$ reduces the MM system to Maxwell EM, the above relation establishes a direct correspondence between the Lorentz-invariant EM field structure and the mean curvature of the manifold. Consequently, equilibrium EM configurations admit a geometric realization through CMC structures. In particular, for closed simply connected manifolds, the equilibrium geometry reduces to the spherical branch established in \cite{Svintradze2025}. Therefore, the Maxwell EM acquires a geometric interpretation through equilibrium CMC manifolds. This reduction, however, does not remove the geometric balance conditions carried by the hypersurface. If $\rho=0$, Eq.~\eqref{eq:gen1} and the tangential balance Eq.~\eqref{eq:gen3} become trivial, but the normal balance Eq.~\eqref{eq:em2} still retains the geometric contribution through $P$, $\sigma B_{\alpha}^{\ \alpha}$, and $\partial_t\sigma$. If an additional static normal limit $C=0$ is enforced, then the normal balance reduces to:
\begin{equation}
P
=
\sigma B_{\alpha}^{\ \alpha}
-
\partial_t\sigma,  \label{CMC1}
\end{equation}
up to the nondegenerate prefactor $\partial_a V^a$. Therefore, the classical Maxwell equations arise from the free Maxwell field condition $\mathcal{F}=0$, whereas field equivalence in the moving-manifold framework remains constrained by the hypersurface's geometry and dynamics.

Having established the EM specialization of the MM system \eqref{eq:em2}--\eqref{CMC1}, we now investigate the conditions under which internal and external electromagnetic configurations become equivalent. The field-equivalence problem is not determined solely by Maxwell electrodynamics, but also by the geometric balance conditions imposed by the moving hypersurface. In particular, Eq.~\eqref{eq:em2} shows that the electromagnetic sector remains coupled to curvature, surface motion, and surface energy through the geometric evolution of $\Sigma(t)$. Consequently, equivalence between internal and external fields can only occur for configurations compatible with the manifold dynamics.

\subsection{Geometric electrodynamics}

Until now, we have examined only the emergence of classical EM phenomena from geometrically evolving boundaries. Here, we focus on how moving boundaries establish field equivalence between an external and an internal manifold bounded by the same hypersurface.  
Let $\underset{\sim}{F}{}^{ab}$ and $\overset{\sim}{F}{}^{ab}$ denote the internal and external electromagnetic field tensors, respectively, and define the field difference by
\[
\Delta F^{ab}
=
\underset{\sim}{F}{}^{ab}
-
\overset{\sim}{F}{}^{ab}.
\]
Applying Eq.~\eqref{eq:maxwell} separately to the internal and external configurations and subtracting the resulting equations yields
\begin{equation}
\partial_{b}\Delta F^{ba}
=
\mu_0 \Delta J^{a},
\label{eq:mismatch}
\end{equation}
where
\(
\Delta J^{a}
=
\underset{\sim}{J}{}^{a}
-
\overset{\sim}{J}{}^{a}
\)
is the difference in current between the two sectors. We focus on closed, isolated configurations for which no interfacial current exchange occurs across the moving hypersurface $\Sigma(t)$. Geometrically, this requires the normal current difference to vanish,
\(
N_a\Delta J^{a}=0,
\)
together with the absence of net sources in the difference sector. Under these conditions,
\(
\Delta J^{a}=0,
\)
and Eq.~\eqref{eq:mismatch} reduces to
\begin{equation}
\partial_{b}\Delta F^{ba}
=
0.
\label{eq:free}
\end{equation}
Equation~\eqref{eq:free} represents a source-free Maxwell system for the field difference. However, within the moving-manifold framework, admissibility of the difference sector is further constrained by the geometric balance condition, Eq.~\eqref{eq:em2}. In particular, static configurations generically retain a nonvanishing geometric contribution via \eqref{CMC1}, so that curvature and surface dynamics continue to produce an effective electromagnetic contrast between the internal and external sectors. In the absence of admissible free electromagnetic modes, such as bounded non-radiative difference configurations, the only admissible solution is
\(
\Delta F^{ab}=0,
\)
which implies
\begin{equation}
\underset{\sim}{F}{}^{ab}
=
\overset{\sim}{F}{}^{ab}.
\label{eq:matching}
\end{equation}
Therefore, according to \eqref{eq:em2}--\eqref{CMC1}, the field equivalence \eqref{eq:mismatch},\eqref{eq:free} emerges not merely as a consequence of Maxwell electrodynamics but as a dynamically constrained geometric regime of the coupled moving-manifold system we examine in detail next.

The MM system naturally separates into static and dynamical regimes. In the static limit
\(
C=0,
\,
V^\alpha=0,
\,
\sigma=\mathrm{const}\neq0,
\)
the normal balance Eq.~\eqref{eq:em2} together with \eqref{CMC-Maxwell}, \eqref{CMC1} reduces to
\begin{equation}
P=\sigma B_\alpha^{\ \alpha},
\label{eq:CMC}
\end{equation}
showing that the effective electromagnetic response is directly induced by the mean curvature of the hypersurface. Consequently, generic static curved configurations retain a nonvanishing electromagnetic contrast and therefore remain visible. In this sense, field equivalence is intrinsically dynamical unless additional geometric constraints are satisfied.

We obtain a broader admissible sector by considering the normal-static regime \(C=0,\, \rho\neq 0\). Then Eq.~\eqref{eq:gen1} reduces to
\begin{equation}
\dot{\nabla}\rho+\nabla_\alpha(\rho V^\alpha)=0,
\label{eq:mass_static}
\end{equation}
which expresses conservation of the surface density along the tangential flow. In this regime, the reduced normal balance \eqref{eq:gen2}, \eqref{eq:em2} takes the form
\begin{equation}
P+\partial_t\sigma-\sigma B_\alpha^{\ \alpha}
-\rho V^\alpha V^\beta B_{\alpha\beta}=0,
\label{eq:nonvacuum_balance}
\end{equation}
where the effective pressure contribution has been absorbed into the definition of \(P\). Hence, according to \eqref{eq:gen3} the tangential dynamics is governed by
\begin{equation}
\rho\left(
\dot{\nabla}V^\alpha
+
V^\beta\nabla_\beta V^\alpha
\right)
=
-\nabla^\alpha\sigma.
\label{eq:tangent_sector}
\end{equation}
Then, the vacuum choice
\(
\rho=0,
\,
\sigma=0,
\,
P=0
\)
gives a trivial solution but does not represent a material bounded configuration. A nonvacuum realization is instead obtained for
\(
\rho\neq0,
\,
\sigma=\mathrm{const}\neq0, \, P\neq 0
\)
for which Eq.~\eqref{eq:tangent_sector} reduces to
\begin{equation}
\dot{\nabla}V^\alpha
+
V^\beta\nabla_\beta V^\alpha
=
0.
\label{eq:flat_flow}
\end{equation}
Equation~\eqref{eq:flat_flow} admits constant tangential velocity fields as simple solutions. In the flat case
\(
B_{\alpha\beta}=0,
\,
C=0,
\,
\sigma=\mathrm{const},
\)
Eq.~\eqref{eq:nonvacuum_balance} determines the effective pressure directly from the tangential motion of the hypersurface. Consequently, the field-equivalence condition possesses explicit nonvacuum solutions with
\[
\rho\neq0,
\qquad
V^\alpha=\mathrm{const},
\]
corresponding to dynamically evolving flat manifolds. In the two-dimensional case, this regime reduces to an infinite planar geometry that supports constant tangential flow while maintaining a vanishing curvature contribution.

More generally, Eq.~\eqref{eq:nonvacuum_balance} shows that a static configuration is not rendered invisible merely by setting the velocity field to zero. Curvature and surface-energy evolution may still generate a nonvanishing effective pressure. In particular, the geometric contribution disappears whenever
\begin{equation}
\sigma B_\alpha^\alpha-\partial_t\sigma=0. \label{time_sigma_balance}
\end{equation}
A special subclass is obtained for
\(
B_\alpha^\alpha=0,
\,
\partial_t\sigma=0,
\)
where the effective pressure vanishes despite the possible presence of nontrivial curvature tensor components \(B_{\alpha\beta}\). This identifies a nontrivial geometric class in which field equivalence may occur without requiring flatness. However, nontrivial compact configurations satisfying
\(
B_\alpha^\alpha=0
\)
do not exist for closed hypersurfaces smoothly embedded in Euclidean space. Consequently, bounded static field-equivalent configurations are generically excluded in Euclidean geometry, whereas dynamically evolving configurations remain admissible through the tangential sector of the moving-manifold system. In higher-dimensional and pseudo-Riemannian settings, the admissible geometric class becomes substantially richer and connects naturally with structures studied in differential geometry and general relativity \cite{ChoquetBruhat2009}.

\section{Discussion and conclusion}

In this work, we formulated electromagnetic field equivalence within the framework of moving-manifold dynamics and showed that invisibility emerges as a geometrically constrained dynamical regime rather than as a purely electromagnetic phenomenon. Starting from a generic tensorially invariant evolution system for moving hypersurfaces, we introduced an electromagnetic specialization via the generalized energy density \eqref{EM_pressure}, which couples electromagnetic fields to curvature, normal motion, tangential flow, and surface energy through the action \eqref{Lagrangian}. Within this formulation, Maxwell electrodynamics naturally arises as a special solution of the broader geometric system.

The analysis of generic equations \eqref{eq:gen1}--\eqref{eq:gen3} and \eqref{eq:em2} shows that field equivalence between internal and external electromagnetic configurations is not determined solely by Maxwell equations. The admissibility of the configuration remains constrained by the geometric balance conditions of the hypersurface. In particular, static configurations generically retain a nonvanishing effective pressure through curvature contributions of the form
\[
P\sim \sigma B_\alpha^\alpha-\partial_t\sigma,
\]
so that curvature induces an intrinsic electromagnetic contrast. Consequently, generic static curved configurations remain visible, while field equivalence emerges predominantly within the dynamical sector of the moving-manifold system. The obtained results further reveal that the admissible class satisfying the field-equivalence condition is nonempty. Flat configurations provide a simple explicit realization, while more general geometries satisfying \eqref{time_sigma_balance} remain admissible candidates even in the presence of nontrivial curvature tensor components. This establishes that invisibility is not necessarily restricted to trivial flat geometries, but may arise from special geometric balance conditions within the coupled manifold dynamics.

For closed hypersurfaces smoothly embedded in Euclidean space, nontrivial compact static configurations satisfying vanishing mean curvature do not exist. Consequently, bounded static field-equivalent configurations are generically excluded in ordinary Euclidean geometry. This restriction, however, does not eliminate the dynamical sector of the moving-manifold system. In particular, the tangential evolution equations admit nontrivial flow solutions even when the normal velocity vanishes, and the coupled geometric balance conditions continue to allow dynamically evolving field-equivalent configurations. Thus, the present formulation does not exclude invisibility in Euclidean geometry, but instead identifies it as an intrinsically dynamical phenomenon arising from the coupled evolution of geometry and fields. The present analysis does not exclude the existence of nontrivial static field-equivalent configurations. Rather, it identifies a constrained admissible class determined by the geometric balance condition \eqref{eq:nonvacuum_balance}. The explicit geometric realization of such static configurations remains an open problem. In higher-dimensional and pseudo-Riemannian settings, the admissible geometric class becomes substantially richer and connects naturally with structures studied in differential geometry and general relativity.

More broadly, the obtained framework is not restricted to electromagnetic fields. Since the governing equations depend only on the generalized energy density $P$, the same geometric mechanism extends naturally to other physical systems formulated within moving-manifold dynamics. This suggests that field equivalence may represent a generic geometric phenomenon associated with evolving hypersurfaces rather than a property unique to electrodynamics.

The present work identifies a geometric mechanism by which electromagnetic fields may propagate across bounded regions without generating observable contrast between internal and external configurations. Rather than relying on engineered constitutive materials or transformation optics, the mechanism emerges directly from the coupled dynamics of curvature, surface evolution, and field interactions. These results provide a geometric perspective on invisibility and establish a direct link between field equivalence and moving-manifold dynamics.

%\section*{Data accessibility}
%No data were generated or analysed in this study.

%\section*{Authors' contributions}
%D.V.S. conceived the work, carried out the analysis, and wrote the manuscript.

%\section*{Competing interests}
%The author declares no competing interests.

%\section*{Declaration of Generative AI and AI-assisted Technologies in the Writing Process}

%During the preparation of this work, ChatGPT, Gemini, and Grammarly were used exclusively for English grammar, fluency, and stylistic refinement. All scientific concepts, equations, derivations, interpretations, and conclusions are the author's original work. The author reviewed and edited all generated text and takes full responsibility for the content of the manuscript.

%\section*{Funding}
%Insert funding information if desired.

%\section*{Acknowledgements}
%Optional.

%\bibliographystyle{unsrt}

\bibliographystyle{apsrev4-2}

\bibliography{Cloakreferences}

\end{document}